\documentstyle[12pt]{article}
\input epsf

\newcommand{\beq}{\begin{equation}}
\newcommand{\eeq}{\end{equation}}
\newcommand{\gsim}{\lower.7ex\hbox{$
\;\stackrel{\textstyle>}{\sim}\;$}}
\newcommand{\lsim}{\lower.7ex\hbox{$
\;\stackrel{\textstyle<}{\sim}\;$}}

\begin{document}
\begin{titlepage}
\renewcommand{\thefootnote}{\fnsymbol{footnote}}

\begin{center} \Large
{\bf Theoretical Physics Institute}\\
{\bf University of Minnesota}
\end{center}
\begin{flushright}
TPI-MINN-95/32-T\\
UMN-TH-1416-95\\
August1995
\end{flushright}
\vspace{.3cm}
\begin{center} \Large
{\bf The Case of $\alpha_s$:  $Z$ versus Low Energies\\
 or
\\
How Nature Prompts us of  New Physics}

\vspace{1cm}

Talk at the  International Symposium on Particle Theory and
Phenomenology \\
(XVIII Kazimierz Meeting on Particle Physics\\
and\\
1995 Madison Phenomenology Symposium)\\
{\em  Iowa State University, Ames, May 22 - 24, 1995}

\end{center}
\vspace*{.3cm}
\begin{center} {\Large
M. Shifman} \\
\vspace{0.4cm}
{\it  Theoretical Physics Institute, Univ. of Minnesota,
Minneapolis, MN 55455}\\
\vspace*{.4cm}

{\Large{\bf Abstract}}
\end{center}

\vspace*{.2cm}
The values of $\alpha_s$ determined from
low- and high-energy measurements
are in irreconcilable contradiction with each other.
The current status of the problem is critically reviewed.
Consequences of the $\alpha_s$ contradiction, in conjunction with
other anomalies detected at the $Z$ peak, are discussed. The
write-up is updated in accordance with  experimental numbers
reported
at summer conferences.

\end{titlepage}

\section{Introduction}

About a year ago I was at the Glasgow Conference. Quite a few talks
there were devoted to the global fits at the $Z$ peak. The message of
all of the speakers,
one after another, was the same:
everything perfectly fits the Standard Model (SM),  no
indications  of new physics are detected.  The conclusion was so
astonishing and so obviously wrong, that shortly after I  wrote
 a ``sociological" paper,
 whose
message
was perfectly opposite \cite{MS}.  If one takes the $Z$ data on
$\alpha_s$ seriously,
one {\em must} insist that new physics is already with
us. The value of the strong coupling constant,
$\alpha_s(M_Z)\approx 0.125$, as it is usually quoted in connection
with the  measurements at the $Z$ peak, is too high to be compatible
with
low-energy phenomenology. It presents a clear signal of the
presence of something
 beyond the Standard Model. This ``something" may very well
be
low-energy supersymmetry, with  sparticle masses in the 100
GeV ballpark, or a more exotic beast.

Today, the psychological climate is totally different. The contradiction
between the low-energy and high energy determinations of
$\alpha_s$ is perceived seriously by many. In several publications
that have
appeared recently \cite{A} -- \cite{B} this contradiction serves as the
starting point for dedicated analyses showing  that the
quality of  global fits at the $Z$ peak  improves if,
instead of the
Standard Model, one considers its supersymmetric generalization,
MSSM.
More than that. Discussion of implications of
low $\alpha_s$ for the general
model-building  started \cite{X} -- \cite{Y}. One of the ideas
is getting information about the
GUT scale corrections, including the masses of the superheavy
particles, which results, in turn, in new and quite specific predictions
concerning
the proton lifetime \cite{X}. Another idea is  relaxing various mass
relations routinely  imposed on MSSM for aesthetical reasons
\cite{Ros,Y}. One then
asks a down-to-earth question: ``what values of the sparticle masses
could explain the observed excess in $\alpha_s (M_Z)$?"  The answer
turns out to be exciting. Let me quote, for instance, Wells and Kane
\cite{Y}: ``... stop and chargino must be light enough to be detected
when the energy of LEP is increased to over 140 GeV ...
If a stop or chargino is not found then either $R_b$ excess
will go away, or if it persists, the SUSY explanation is not relevant..."
Here, the authors mention also, another puzzle of the $Z$ physics,
$R_b$, to be discuss in brief later on.
The so called $R_b$ crisis which surfaced recently tends to eclipse
the confrontation between the low- and high-energy values of
$\alpha_s$. It should be stressed, however, that these two problems
are totally unrelated to each other experimentally since the
methods used for measuring $R_b$ and $\alpha_s$
at the $Z$ peak are completely different. Even if one assumes, for a
short while,  that the $R_b$ excess is a spurious instrumental effect,
this
need not be the case with $\alpha_s$. At the very least, one should
say that there are two independent measurements signaling
deficiencies of the Standard model at the $Z$ peak, so that the
probabilities that these two effects are statistical fluctuations are
multiplied.
Thus, it is
impossible to overestimate the lead provided by $\alpha_s
(M_Z)$, especially in conjunction with the $R_b$
crisis,
the only new hints we've gotten from Nature over a decade.

My talk  consists of three parts. First, I will argue that the
genuine
strong coupling constant at $Z$ is close to 0.11. Second, I will briefly
discuss implications of this fact for physics at the 100  GeV energy
scale. At the end I will comment on this problem in the context of
Grand Unification.

\section{Large versus small $\alpha_s (M_Z)$}

When I say small or large  $\alpha_s $, I mean the following:
if $\alpha_s (M_Z)$ is close to 0.11 it will be referred to as small;
if it is close to 0.125 it is large ($\alpha_s$ is defined in the
$\overline{\mbox{MS}}$ scheme \cite{MSbar}). The distinction
between these two
values is clearly seen on
Fig. 1 which presents a compilation of data on different
measurements
of $\alpha_s$, essentially borrowed  from Bethke's talk \cite{Bethke},
with a few points updated.  First,
I  erased a
couple of points with  error bars too large  to be informative. They
merely overshadow the general picture. Added is a point
obtained recently by Voloshin \cite{Volo} from analysis of the QCD
sum rules in the $b\bar b$ channel. The result claims to have error
bars so
small,  they are barely seen on the plot. I hasten to add
though, that the estimate
of the uncertainties presented in Ref. \cite{Volo} seems to refer to
the uncertainty of a particular procedure, and should be taken with
caution. Finally, the lattice prediction for $\alpha_s$ is quoted from
the recent talk \cite{CM}.

The pattern is quite obvious: all low-energy
measurements cluster around a small value of the strong coupling
constant, $\alpha_s(M_Z)\approx 0.11$, with one notable exception
of $\tau$ decays, to be discussed later.  Determinations of $\alpha_s$
that are the cleanest
from the
theoretical standpoint are those done in the Euclidean domain --
deep inelastic scattering, lattices, and the Voloshin sum rule.
The corresponding  four points are marked by the
arrows.

At the same time, the conventional routine of determining
$\alpha_s$ at the
$Z$ peak under the SM assumptions leads to high values,
clustering around $0.125$.

The $\tau$ decays  will be  subject to
special scrutiny below, and you will hopefully see that the theoretical
uncertainty usually quoted is  grossly underestimated. If a proper
value of this uncertainty is used, the $\tau$ point must be merely
ignored as uninformative. The drawback of the $\tau$ analysis
is its essentially Minkowskean nature. Nonperturbative contributions
in the Minkowski domain are expected to be essentially larger (and
die off
much slower) than similar contributions ``from the other side",
in the Euclidean domain. Moreover, it is very difficult (if possible
at all) to control them in the Minkowski domain based
on the truncated condensate series.

Taken at its face value,    the discrepancy between the small and
large
$\alpha_s$ clusters might seem quite marginal -- this is a two
$\sigma$ effect or so,
and who cares about
2$\sigma$ effects?  Being expressed in terms of the low-energy
parameters the difference becomes pronounced, however. Indeed,
the
large value of $\alpha_s(M_Z) =0.125$
is translated in $\Lambda^{(4)}_{\overline{\rm MS}}\approx
$ 480 MeV, to be compared with $\Lambda^{(4)}_{\overline{\rm
MS}}\approx
$ 200 MeV appearing in the small-$\alpha_s$ case. Although
routinely used,  $\Lambda^{(4)}_{\overline{\rm MS}}$ is
far from being a perfect parametrization. (As a matter of fact,
the $\overline{\rm MS}$ scheme as a whole is rather unphysical,
but this is a subject for another story. It is commonly used, by
convention, and I will follow this convention too for the time being.)
At the very least,
it makes more sense to compare $\Lambda^{(3)}_{\overline{\rm
MS}}$ -- this parameter is more relevant to
the low-energy hadronic phenomenology and is somewhat  larger
than $\Lambda^{(4)}_{\overline{\rm MS}}$. Anyway, the situation is
quite transparent. From the point of view of the low-energy QCD the
contradiction between the first and the second values of
$\Lambda^{(4)}_{\overline{\rm MS}}$
 is qualitative and irreconcilable. If the scale parameter of QCD is so
high
one must be prepared to say farewell to a whole wealth of results
accumulated in QCD over years.  The success of  QCD sum rules in
dozens of
problems referring to all aspects of  hadronic phenomenology
must be considered as a pure coincidence then, a conclusion which
seems quite fantastic.

To illustrate my point let me show you a typical plot one deals with
in
the
analysis of the sum rules. (Fig. 2).  The plot presents the
Borel-transformed correlation
function
$\Pi$
of two vector currents (with the isotopic spin $I=1$),
$J_\mu = (1/2)(\bar u\gamma_\mu u -\bar d\gamma_\mu d)$,
versus $M$, the (Euclidean) Borel parameter. Figure 2 shows the
integral
\begin{equation}
I(M)  = \frac{2}{3M^2}\int ds R^{I=1}(s) e^{-s/M^2} \, .
\label{SR}
\end{equation}
The imaginary part
of this correlation function in the Minkowski domain is known, and
is proportional to the total cross section of the $e^+e^-$ annihilation
into
hadrons
with the isotopic spin $I$ equal to 1. This cross section is measured
with a
high precision near the $\rho$ meson and below. The accuracy gets
somewhat worse at energies
above 1.5 GeV, but this fact is irrelevant for two reasons: (i) at
$M<1.1$
GeV
the $\rho$ meson contribution exceeds 70\%; (ii) it is the absolute
normalization of the individual measurements of $R^{I=1}(s)$ which
is most uncertain.
This means that  the genuine experimental curve runs  inside the
corridor labeled by ``U' and ``L" on Fig. 2; the shape of the curve is
``parallel" to the upper and lower sides of the corridor (``U' and ``L").
In other words
one can slightly shift the curve for
$I(M)$ upwards or downwards, as a whole, without distorting the
shape of the curve. It is practically impossible to deform the {\em
form} of the
curve since it results from integration over a large number of
points.

The experimental corridor I use was obtained in Ref. \cite{EKV}
in the following way. Each individual point in the set of all
measurements of $R^{I=1}(s)$ has a systematic uncertainty. If the
uncertainties in {\em all} individual points are taken with the plus
sign we get the upper side (``U"), if they are taken with the minus
sign we get the lower side (``L"). It is clear that a more careful
analysis could significantly narrow the corridor, the more so that
several new points obtained since 1979 have smaller error bars.
But even the old data seem to be sufficient to make a definite
conclusion.

The most remarkable thing is not what we see on the plot but,
rather, what
we do not see there. Namely, if we {\em descend} from higher values
of $M$ to lower ones, within the window, the experimental curve
shows no indication whatsoever of growth or explosion of $I(M)$. On
the
contrary, $I(M)$ stays practically constant (although this constant
value is somewhat uncertain), and then smoothly goes down at
$M\sim$ 1.0 GeV, i.e. slightly above the $\rho$ meson mass.
If we believe that the description based on perturbation theory plus
a few nonperturbative condensates is applicable in the window
shown on Fig. 2 we have to conclude that $\alpha_s$ is small.
The curved labeled by ``220" on Fig. 2 corresponds to
$\Lambda_{\overline{\rm MS}}^{(3)}= 220$ MeV and the standard
values of the gluon and four-quark condensates. The shaded area
indicates my estimate of the theoretical uncertainty due to
$\alpha_s^3$ terms and higher condensates (the uncertainty in the
gluon and four-quark condensate is not included).  It is clearly seen
that this theoretical curve is reasonable.
At the same time, if
$\Lambda_{\overline{\rm MS}}^{(3)}$ is
in the vicinity of 550 MeV (this value follows from
$\alpha_s (M_Z) =0.125$) the strong coupling constant calculated
as a function of $M$  explodes at $\sim$ 1.5 GeV.
As a reflection of this explosion the theoretical curve goes
up sharply, as shown on Fig. 2. Even $\Lambda_{\overline{\rm
MS}}^{(3)} = 400$ MeV is hardly acceptable. To give an idea of
the role of the condensates I plot two curves corresponding to
$\Lambda_{\overline{\rm MS}}^{(3)} = 400$ MeV --
one with the standard value of the four-quark condensate, and the
other one (labeled by ``2 QC") with twice the standard value.

The
difference is striking and qualitative. If $\Lambda_{\overline{\rm
MS}}^{(3)} < 250$ MeV the explosion occurs far below the window,
so that no reflection of the explosion is seen
in the theoretical curve.  The pattern of behavior in the transitional
domain is decided mostly by the condensate corrections. For large
values of $\Lambda_{\overline{\rm
MS}}^{(3)}$ we  are unable to  descend to the domain where
the condensate corrections play a role. The explosion of the
perturbative series occurs earlier.

I emphasize  that this situation is typical for
the QCD sum rules. That is why I concluded that
they can not be reconciled with the high value of $\alpha_s$.
Needless to say this is not a mathematical theorem. Being stubborn,
one could always insist that
 the flatness of $I(M)$ could be a
result of a general
conspiracy -- $\Lambda_{\overline{\rm MS}}^{(3)}$ is very large,
terms
of all orders in $\alpha_s$ are large at $M\sim $ 1 GeV but  they
combine
 to produce a smooth curve. All successful predictions of
sum rules are mere coincidence.
As usual, the theory of general conspiracy is impossible to rule out
by rational arguments.

\subsection{The $Z$ peak kitchen.}

At least four experimental groups have extracted $\alpha_s$, by
doing different
measurements at $Z$ (see e.g. \cite{zed,Olch}). The most precise
approach, both theoretical and experimental, is the
measurement of the total hadronic width of the $Z$. Not only is this
method  most straightforward experimentally, the theoretical
uncertainty here is the smallest since (i) this is the only quantity
calculated to the order ${\cal O}(\alpha_s^3)$; (ii) non-perturbative
corrections can be argued to be minimal in the total width. In
the $\overline{\rm MS}$ renormalization scheme, and assuming the
validity of the Standard Model,
the result
for the total hadronic $Z$ width $\Gamma_h$ has the form
\begin{equation}
\Gamma_h = \Gamma_0 \left( 1 +r_1\frac{\alpha_s (M_Z)}{\pi}
+r_2\frac{\alpha_s (M_Z)^2}{\pi^2} +r_3\frac{\alpha_s (M_Z)^3}{\pi^3}
\right)
\label{gammah}
\end{equation}
where $\Gamma_0$ is the parton model width
(with the electroweak corrections included), and the coefficients
$r_1,r_2$ and $r_3$ are obtained by computing  relevant
Feynman graphs, see  \cite{gammaz,kataev} and references therein.
Nice reviews are also given in Refs. \cite{Chet,SS}.
The coefficients depend on the quark masses; the most noticeable
uncertainty of this type is due to the errors in the $t$ quark mass.
The coefficient $r_1$ is unity (modulo tiny mass corrections),
$r_2$ is of order unity, $r_3$ is negative and is of order 10.
This pattern suggests that at order $\alpha_s^4$ the asymptotic
nature of the $\alpha_s$ series will, perhaps, show up so that
including the ${\cal O}(\alpha_s^4)$ term is not going
to improve the theoretical accuracy \cite{I3}.  At the moment the
experimental uncertainty in this particular quantity,
$ \Gamma_h$, dominates so that there is no point in discussing other
sources of uncertainty. Comparing Eq. (\ref{gammah}) with
$$
\Gamma_h^{exp} = 1744.8\pm 3.0\,\,\,  \mbox{MeV}
$$
one arrives at
\begin{equation}
\alpha_s (M_Z) = 0.125\pm 0.005 \, ,
\end{equation}
see Ref. \cite{Olch}.  This number assumes that there are no
contributions
beyond the Standard Model.

As I have already mentioned, there are good reasons
to believe that nonperturbative effects in $\Gamma_h$ are
negligible since the energy release is sufficiently high.

Notice that even the leading (first) order term in $\alpha_s$ presents
a small correction ($\sim 4\%$).  A 13\% shift in $\alpha_s$
amounts to 5 permille effect in $\Gamma_h$. In absolute numbers
this is of order  10 MeV.

Here we come to a remarkable observation: the experimentally
measured yield in the $b\bar b$ channel is
higher
than the SM expectations by $11\pm 3$ MeV (the $R_b$
crisis). Is this a mere coincidence? It is tempting to say no.
Assuming,
that these extra 11 MeV in the $b\bar b$ width are due to new
physics
and subtracting them from $\Gamma_h$ we  get \cite{Olch}
$$
\alpha_s (M_Z) = 0.102\pm 0.008\, ,
$$
in perfect agreement with the low-energy determinations. This fact
was
mentioned already in 1993 by Altarelli {\em et al.} \cite{Alt} but
was
largely ignored and forgotten. The Glasgow Conference rapporteurs
mentioned in
passing $R_b$ as the only  little light cloud on the face of the
Standard
Model, hastily adding that the signal is statistically insignificant
\cite{webber}.
Now, of course, people are more inclined to speculate on the impact
of the
$R_b$ excess. The situation remains fogged, however,  since, in
addition to the
$R_b$ excess, there is a hint in the data on the deficit in the $\bar
c c $ yield,
so that the excess in the $\bar b b $ yield seems to be offset.
Moreover, this shortage finds no natural explanation in the minimal
supersymmetric extension of the Standard Model \cite{A,3,Y,Olch}
casting shadow on all data referring to the heavy quark yields at $Z$.
The
experimental uncertainty in $R_c$ is  larger than that in $R_b$,
though,
which prevents  unambiguous conclusions at present.
Once again, I suggest to use the
$\alpha_s$ information to make a {\em prediction}:
when the dust is settled an overall excess in $\Gamma_h$  in the
ballpark of 10 MeV will be confirmed.

\subsection{Survey of low-energy determinations of $\alpha_s$}

Many compilations of the low-energy determination of $\alpha_s$
share a common drawback: they are plagued by ``no discrimination
policy". Those who compile the estimates forget
(or do not pay attention) that QCD is a peculiar theory
whose infrared behavior remains unsolved, and various approaches
enjoy different degree of control over the large distance dynamics.
On the other hand,
all experimentally measurable quantities  do involve infrared
dynamics. Even those which are claimed to be infrared stable
are typically protected from the large distance contributions only
in perturbation theory. Therefore, not all low-energy evaluations
$\alpha_s$ are equal to each other.  In particular, I will discard, from
the very beginning, all analyses of the jet parameters (thrust and so
on). Theoretical formulae one uses to extract $\alpha_s$ from these
measurements
are purely perturbative. No reliable methods were suggested so far
allowing one to estimate, even roughly, non-perturbative corrections
since these
processes are not amenable to operator product expansion (OPE)
\cite{OPE}. Moreover, if in the processes tractable within OPE
the nonperturbative corrections start from quadratic terms $1/Q^2$
or $1/E^2$ in the jet physics  (and in the hard  processes
without OPE in general) linear $1/Q$  or $1/E$ nonperturbative
corrections are argued to be quite abundant
\cite{1/Q}. Attempts to build phenomenology of the $1/Q$ terms
based on renormalons and related ideas are in its infancy.
Even the question of their universality is being debated at the
moment, to say nothing about reliable estimates of the coefficients
in front of $1/Q$. Some estimates existing in the literature \cite{1/Q}
indicate that the $1/Q$ corrections can be so large that they
invalidate all existing determinations of $\alpha_s$
from the jet physics. The same criticism applies to the low-$x$
physics at HERA.

Thus, we will consider only those analyses where at least some
control over the nonperturbative effects exists at the present level of
understanding of QCD. This  leaves us with deep inelastic
scattering (DIS) \cite{DIS}, including the Gross-Llewellyn-Smith and
similar sum rules, inclusive decay rates of the type
$\Upsilon\rightarrow
$ photon + hadrons or $\tau\rightarrow \nu$ + hadrons, QCD sum
rules and, finally, lattices. Each of these approaches has its
advantages and drawbacks, and we will briefly discuss them in turn.
{}From the theoretical point of view, deep inelastic scattering and the
QCD sum rules are the ``cleanest"
processes, at least in principle, since  they allow control over
each and every aspect of calculation. The inclusive widths are
essentially Minkowskean quantities; as we will see shortly,
this brings in an additional theoretical uncertainty which is hard to
estimate. Finally, the lattice calculations are burdened by systematic
errors,
associated with  finite size effects and, especially,
putting chiral dynamical quarks on the lattice, whose understanding
is not yet fully settled.

\vspace{0.3cm}
\begin{center}
{\it  (i)  Deep inelastic scattering}
\end{center}
\vspace{0.2cm}

Data on deep inelastic scattering are very abundant. Many high
statistics experiments were analyzed with the aim of extracting
$\alpha_s$ and testing the QCD evolution. QCD predicts not the
structure function themselves but, rather,  evolution with
$Q^2$ increasing. Nonperturbative effects are represented
by higher twists; in the Euclidean domain, where the analysis
is carried out, practically, it is quite sufficient to limit oneself to
twists two and four.  To illustrate the subtle points of the analysis
let us turn to a particular work \cite{VM}. At the very end, I will
quote the world average for $\alpha_s$ from DIS.

Figure 3a, borrowed from Ref. \cite{VM}, shows the $Q^2$ evolution
of the structure function $F_2$ for different values of $x$.
The experimental points are fitted according to the predictions of
QCD, in the next-to-leading logarithmic approximation (NLO), i.e. at
two loops. The dashed line visualizes the $Q^2$ evolution without the
higher-twist effects.  It is seen that at $x$ lying in the interval
0.2  to 0.5 the power corrections are sufficiently small. At higher
values of
$x$ the power corrections increase, on the one hand, and the quality
of the data becomes worse, on the other. Lower values of $x$ are
more sensitive to the gluon distributions, which may bring in
unwanted model dependence. Thus, the above interval of $x$ is
optimal. To avoid contamination from higher twists,
determination of $\alpha_s$ has to be performed  in  the
``high-$Q^2$" domain where the power corrections are negligible.
Depending on the particular value of $x$ chosen, this domain
stretches above 5 to 10 GeV$^2$. In this domain the logarithmic
derivatives $d\ln F_2 /d\ln Q^2$ are very nearly proportional to
$\alpha_s (Q^2)$, with an $x$ dependent proportionality coefficient
that depends only weakly on the $x$ dependence of the measured
$F_2$. Comparing the  measured logarithmic derivatives with
those obtained in QCD in terms of $\alpha_s (Q)$ one fits the
strong coupling constant (see Fig. 3b).  Neither the uncertainty
in the gluon distribution nor the higher twist corrections are
important provided one limits oneself to the interval $0.2<x<0.5$.

The theoretical uncertainties are due to the (virtual) heavy flavor
thresholds and due to the scale ambiguity. The scale ambiguity
emerges because we truncate the perturbative prediction
at the next-to-leading order. Therefore, at this order we do not know
exactly the argument of the quantities involved; it may be $Q^2$
 or $\mu^2 = kQ^2$ where $k$ is a number of order one. Only the
next-to-next-to-leading order calculation fixes $k$ at the
next-to-leading order. Practically, the scale ambiguity is the largest
theoretical uncertainty. Figure 4 shows the sensitivity of the
$\alpha_s$ determinations to scale parameters introduced in a
certain way.  In Ref. \cite{VM} it is suggested to vary $k$ between,
say, one quarter and four.  This would correspond to theoretical
uncertainty $\pm 0.004$ in $\alpha_s(M_Z)$. From  experience
accumulated in the last few years (for instance, from the BLM scale
setting procedure
\cite{BLM}) we know, however, that it is extremely unlikely that
$k>1$. In all problems treated in the literature $k$ turns out to be
$< 1$, a result perfectly natural on  physical grounds since typically
the momentum $Q$ is shared between several quarks and/or gluons.
If so, the theoretical uncertainty should be taken with the minus
sign only.  The fitted value of $\alpha_s (M_Z)$ in Ref. \cite{VM}
is 0.113; experimental and theoretical errors are 0.003 and 0.004,
respectively. Figure 4 demonstrates, to my mind, very
convincingly, that $\alpha_s (M_Z)=0.125$
is way beyond what is allowed by the DIS data.

More recently,  the $Q^2$ evolution of the non-singlet structure
function at high $Q^2$ (i.e. $Q^2> 150\,\, \mbox{GeV}^2$)
was analyzed by the CCFR Collaboration \cite{CCDIS}. Their result is
$$
\alpha_s (M_Z)= 0.111\pm 0.002 \; \mbox{(stat)}\;\pm
0.003\; \mbox{(syst)}\, .
$$

The fresh world average of the $\alpha_s$ determinations from the
neutrino and muon deep inelastic scattering is presented in Ref.
\cite{Eisele}, which gives practically the same number,
\begin{equation}
\alpha_s (M_Z)_{DIS}=0.112\pm 0.005\, .
\end{equation}

\vspace{0.3cm}
\begin{center}
{\it  (ii)  The Gross-Llewellyn Smith (GLS) sum rule}
\end{center}
\vspace{0.2cm}

This sum rule predicts the value of the integral
\begin{equation}
\int_0^1 F_3^{NS} (x, Q^2) dx = 3\left\{ 1 -\frac{\alpha_s (Q)}{\pi} + ...
+ \,\, 1/Q^2 \,\, \mbox{corrections}\right\}\, .
\end{equation}
In the asymptotic limit $Q^2\rightarrow\infty$, the right-hand side
tends to 3. If we make $Q^2$ sufficiently large so that
the higher-twist effects are already unimportant deviations from 3
fix the value of $\alpha_s$.  A state-of-the-art theoretical description
of the Gross-Llewellyn-Smith sum rule was presented recently
\cite{CK}, including the next-to-next-to leading order perturbative
corrections \cite{I1} and higher twist effects \cite{I2}. This work
triggered a new data
analysis by the CCFR Collaboration; the experimental values
for the GLS integral versus $Q^2$ were updated. I give here the plot
(Fig. 5) borrowed from Ref. \cite{Harris} (see also the review talk
\cite{Eisele}).
The solid line on this plot corresponds to a QCD fit with
$\Lambda^{(5)}_{\overline{\rm
MS}} = 150$ MeV (or $\alpha_s (M_Z) = 0.112$). It is important that
the $\alpha_s$ corrections are
negative.  Therefore, with $\Lambda$ increasing  the fit curve would
shift further down, in clear
conflict
with the data plotted on Fig. 5.
The outcome of the
analysis is
\begin{equation}
\alpha_s (M_Z)_{GLS}=0.108\; ^{+0.003}_{-0.005}\;
\mbox{(stat)}\,\pm 0.004\;  \mbox{(syst)}\; ^{ +0.004}_{-0.006}\;
\mbox{(higher twist)}\, .
\end{equation}

\vspace{0.3cm}
\begin{center}
{\it  (iii)  The QCD sum rules}
\end{center}
\vspace{0.2cm}

In the beginning of my talk I  presented a plot
visualizing a typical sum rule in the classical $\rho$ meson channel
(Fig. 2). A few explanatory remarks are in order here concerning
both the
theoretical and experimental sides of the sum rule.  For a general
introduction to the method, see Ref. \cite{shif}.

The right-hand side of the sum rule (\ref{SR}) is, in principle,
measurable. In practice the cross section of the $e^+e^-$ annihilation
to hadrons with the unit total isotopic spin is rather poorly known
at $E=\sqrt{s}>2$ GeV.  If $M$ is sufficiently small, however,
the exponential weight in Eq. (\ref{SR}) damps the high-energy tail
of the integral.  Even if the integrand is known in this domain
with 10\% accuracy (integrally), which seems to be more
than realistic,  the impact on the uncertainty of
the sum rule will
be at the level of a fraction of one percent provided $M\sim$ 1 GeV,
i.e. totally negligible for our purposes. Therefore, we can use the fact
that at $E>2$ GeV, the measured value of $R^{I=1}(s)$ must coincide
with the one calculated perturbatively with the accuracy better than
10\%.  We then glue a theoretical tail to the experimental curve
keeping in mind that the left-hand side is very insensitive to
variations of the tail within the reasonable limits.

The perturbative calculation of $R^{I=1}(s)$ is carried out at three
loops \cite{3L}; with three active flavors we have
\begin{equation}
R^{I=1}(s) = \frac{3}{2} \left\{ 1
+\frac{\alpha_s (s)}{\pi} + 1.64 \left( \frac{\alpha_s (s)}{\pi}\right)^2
-10.2 \left( \frac{\alpha_s (s)}{\pi}\right)^3 +...\right\} \, .
\label{Rpert}
\end{equation}
In the domain of interest $\alpha_s /\pi$ is in the ballpark of 0.1.
A glance at Eq. (\ref{Rpert}) leads us to conclude that
the third term  is of the order of the second one, and the
truncation of the perturbative series is near a critical point. We will
return to this issue later on. Here I note that under
the
circumstances, including the third term in the analysis does not
improve the accuracy of the theoretical prediction. Rather, we should
look at it as a natural measure of the maximal theoretical accuracy
one can achieve.  The relative strength of the third term is
close to 1\%; this means that any analysis in this energy range
can measure $\alpha_s$ with the 10\% uncertainty, at best,
which translates into 3\% uncertainty in $\alpha_s (M_Z)$.
This is not bad, indeed, since the difference between
the large and small $\alpha_s (M_Z)$ is around 13\%.

With this understanding in mind we proceed to comparing the
theoretical and experimental sides of the sum rule (\ref{SR}).
Let us start from the theoretical side known  since almost
prehistoric times,
$$
I(M) = 1 + \frac{\alpha_s(M)}{\pi} + 2.94
\left(\frac{\alpha_s(M)}{\pi}\right)^2 + {\cal O}
\left((\alpha_s/\pi )^3\right) +
$$
\begin{equation}
+C_G \langle {\cal O}_G\rangle \frac{1}{M^4}
+C_q \langle {\cal O}_q\rangle \frac{1}{M^6}
+ {\cal O} (M^{-8}) \, .
\label{TSSR}
\end{equation}
Here ${\cal O}_G$ and ${\cal O}_q$ are the gluon and the four-quark
condensates, respectively \cite{shif}, and  $C_G$ and $C_q$ are their
coefficients. A few comments are in order here concerning the
perturbative and nonperturbative parts represented by the first and
the second lines in Eq. (\ref{TSSR}).

Notice that the $\alpha_s^2$ coefficients in Eqs. (\ref{Rpert})
and (\ref{TSSR}) are different.  The reason for this is the Borel
transformation used to obtain the sum rule (\ref{SR}). It is not
difficult to show that under this transformation
$\alpha_s /\pi$ in $R$ (and in the Adler $D$ function)
goes into
$$
 \frac{\alpha_s(M)}{\pi}  +\frac{9{\bf C}}{4}
\left(\frac{\alpha_s(M)}{\pi}\right)^2 + \mbox{higher orders}
$$
where ${\bf C}$ is the Euler constant.

The perturbative expansion in
Eq. (\ref{TSSR}) must be supplemented by the running formula
for $\alpha_s$, which for three active (massless) quarks
takes the form \cite{SSA}
\begin{equation}
\frac{\alpha_s (M)}{\pi}
= \frac{4}{9} \left(\ln\frac{M^2}{\Lambda^2}\right)^{-1}
-
\frac{256}{729}\left( \ln\ln\frac{M^2}{\Lambda^2}\right) \left(\ln
\frac{M^2}{\Lambda^2}\right)^{-2} +...
\label{RA}
\end{equation}
In the $\alpha_s^2$ term in Eq. (\ref{TSSR}) it is legitimate
to substitute the one-loop formula for $\alpha_s$ since the
difference is of the  higher  order. (To make arithmetics simpler I
used, however, the two-loop expression for $\alpha_s$ everywhere.)
The expansion (\ref{RA}) is good only if $\ln (M/\Lambda )^2 \gg 1$.
In our window this logarithm is not large, especially for higher
values of $\Lambda$.   Equation (\ref{RA}) is used literally, however;
as an educated guess for the uncertainty I took the $\alpha_s^3$
term in $I(M)$ and in the running law of the strong coupling
constant.

In the domain of $M$  where we are going to work
the logarithm $\ln  (M^2/\Lambda^2)$ is such that
the two-loop terms in $I(M)$ and $\alpha_s (M)$ nearly compensate
each other. Still the positive term $2.94 (\alpha_s /\pi )^2$
in $I(M)$ is somewhat larger than the negative $1/\ln^2$
contribution coming from $\alpha_s /\pi $. This means that
the overall $1/\ln^2$ contribution to $I(M)$ is rather small and
{\em positive}.  The positivity implies that the fit of the sum rule
with the two-loop accuracy will necessary produce a {\em lower}
bound on the value
of $\Lambda$ than the fit performed with the one-loop formulae.

Let us turn now to the nonperturbative terms. The gluon condensate
is defined as
$$
\langle {\cal O}_G \rangle = \langle\frac{\alpha_s}{\pi}
G_{\mu\nu}^aG_{\mu\nu}^a
\rangle \, .
$$
The coefficient $C_G$ was calculated at one-loop order in \cite{shif},
$$
C_G = \frac{\pi^2}{3}\, .
$$
The two-loop answer is also known \cite{sursa}. We will not need
this two-loop result
however. The correction is quite modest and is by far smaller
than the existing uncertainty in the numerical value of the gluon
condensate.

The four-quark condensate has a generic structure
$$
{\cal O}_q = \bar q\Gamma q \bar q\Gamma q
$$
where $\Gamma$ stands for a combination of the Lorentz and
color matrices.  Two different combinations are relevant in the vector
channel. I will not go into details since they are described at length
in Ref. \cite{shif}.  Anyway, the vacuum matrix element of ${\cal
O}_q$
is known only within the factorization hypothesis. Factorization
becomes exact in the large $N_c$ limit. Certainly, at $N_c=3$
one could expect some deviations from factorization.  For those
operators which appear in the vector channel,
the deviations are small, however. This fact was checked
 within the
sum rules themselves, see e.g.  \cite{DFS}, and on the lattices
\cite{mart}. It seems quite safe to say that possible deviations from
factorization are smaller than the uncertainty in $\langle\bar
qq\rangle$. The anomalous dimension of the operator ${\cal O}_q$
is such, that it practically compensates the logarithmic dependence of
$\alpha_s (M)$ appearing in $C_q$. Assembling all these elements
together one gets \cite{shif}
$$
C_q \langle {\cal O}_q \rangle =
-\frac{448}{81}\alpha_s (\mu ) \langle \bar qq (\mu )\rangle^2
$$
where $\mu$ is a low normalization point, of the order of
the typical hadronic scale, say $\sim 700$ MeV.  If one uses the
``standard" numerical values of the gluon and quark condensates
the nonperturbative part of $I(M)$ takes the form
\begin{equation}
I(M)_{np} = 1 + 0.1 \left( \frac{0.6}{M^2}\right)^2 - 0.14
\left( \frac{0.6}{M^2}\right)^3
\label{SRNP}
\end{equation}
where $M$ is measured in GeV, the term $M^{-4}$ is due to the gluon
condensate, while the last term is due to the quark condensate. By
the standard values I mean those accepted in Ref. \cite{shif}.
The uncertainty in the gluon condensate is $\sim 30 \%$ while in the
four-quark condensate
it can be as large as, perhaps,  factor of 2.

If the gauge coupling
constant is
set to zero and quarks are allowed to propagate freely, then
$I(M) = 1$.
With the interaction switched on $I(M)$ deviates from unity,
but the deviation is not large in the $M$ interval shown.

The theoretical and experimental sides of the sum rule are compared
on Fig. 2 which has been already discussed previously.

I pause here to make a remark concerning the truncation of the
perturbative and power series in the theoretical side of the sum rule.
As well-known, both series are asymptotic. The asymptotic nature of
the $\alpha_s$ expansion in QCD was established long ago
\cite{thooft}. Recently it was proven \cite{pascos} that the
condensate series is also asymptotic. This means that including
more and more terms of  expansion in the theoretical prediction
does not necessarily improves its accuracy. Since $\alpha_s $ is
rather
large in the domain of interest, $\alpha_s \sim 0.3$,  it is clear that
we must limit ourselves to  just a few terms.  A naive examination of
Eq. (\ref{Rpert}) tells us that already the third term seems to fall into
the tail
of the asymptotic series which must be discarded. There is a fancy
way to come  to the same conclusion:  summation of the so
called renormalon chains. The renormalon chain is a sum of bubble
contributions of any order in $\alpha_s$ presenting a subset of
factorially divergent Feynman graphs. Since this is a very specific
subset of all possible diagrams, I do not think that the renormalon
chains are useful in the quantitative sense. The narrow choice of the
set of graphs is not the main reason, however. The
renormalons are inconsistent with the OPE-based approach, which is
the only known basis for the proper treatment of QCD. Therefore, I
would not say that including them  improves our accuracy. They may
serve, however,  as an  estimate of the uncertainty one can expect
from theoretical formulae, in the absence of better ideas.  Analysis of
the renormalon chains in the context we are interested in, was
carried out in Refs. \cite{R1} -- \cite{R3} (as a matter of fact, some of
these
works were mainly devoted to the $\tau$ decays; the spectral
densities in both cases are similar, and the conclusions are applicable
to the
sum rule (\ref{SR}) as well).  And, sure enough, the outcome of
this rather sophisticated analysis boils down to saying that the
uncertainty is of the order of the third term in Eq. (\ref{Rpert}).

For the same reasons inclusion of the power terms of high dimension
seems to make no  sense.

\vspace{0.3cm}
\begin{center}
{\it  (iv)  Sum rules in the $b\bar b$ channel}
\end{center}
\vspace{0.2cm}

Voloshin's analysis \cite{Volo}, as it stands now, gives the most
accurate evaluation of $\alpha_s$, namely, $\alpha_s (M_Z)
=0.109\pm 0.001$.  The two-point function in the
$b\bar b$ channel is calculated.  The $b$ quarks are treated in
the nonrelativistic approximation.  I would say that the work is very
close in spirit to one of the recent lattice calculations \cite{NRQCD} --
the only distinction is that on the lattices the correlation function is
calculated numerically (in order to set the scale in the physical
units), while here we deal with an analytic calculation. Both
approaches rely on experimental data in the $b\bar b$ channel.
What is remarkable,  is that the problems and difficulties one
encounters
are close in  nature. For instance,
higher-order relativistic corrections are disregarded in both cases.
The same is valid for higher-order $\alpha_s$ corrections.
Of course, the lattice analysis has its specific problems (especially
with the light quarks) which are absent in  analytic QCD.
Since the stakes are very high it seems reasonable to be
conservative. Making various {\em extreme} assumptions about
possible effects due to higher-order relativistic and $\alpha_s$
corrections
not accounted for in the sum rules  \cite{Volo} will probably allow
one to stretch the error bars up to 0.003 or 0.004.

\vspace{0.3cm}
\begin{center}
{\it  (iv)  The lattice calculation}
\end{center}
\vspace{0.2cm}

To complete the list of the  Euclidean approaches  used
for determination of  $\Lambda$ or $\alpha_s$ let me mention the
lattice calculations. The idea is straightforward -- one fits the
calculated
spectrum in the $b\bar b$ or $c\bar c$ channels to the experimental
one
fixing in this way $\Lambda_{\rm lattice}$ in the physical units.
Certainly, one has to overcome many technical problems and
conceptual difficulties, especially with the dynamical light quarks.
Having only quite a superficial idea about the lattice calculations
I can not seriously comment on that. Fortunately, there is no need,
since we had two beautiful talks at this Symposium
\cite{Aida,Lepa}.  Details can be inferred from these talks. As far as I
understand the main limitation is due to the fact that the full QCD
lattice simulations have been carried out on a rather coarse lattice. In
this sense the recent lattice revolution
\cite{LR} turns out to be quite helpful. For instance, the NRQCD
Collaboration uses the effective Lagrangian method to improve the
coarse lattice spacing and extrapolates the results referring
to $N_f =0$ and $N_f=2$ to $N_f = 3$. Using the one-loop
perturbative matching for the plaquette action this group obtains
\cite{NRQCD}
$\alpha_s(M_Z) = 0.115\pm 0.003$. I suspect that the quoted error
does not fully reflect the systematic uncertainties of the method.
More specifically, I suspect that the one-loop expression which
connects the measured plaquette action with the gauge coupling
constant may be less accurate than  quoted above. If,
say, the two-loop contribution is twice  the square of the
one-loop
(and this does not seem crazy to me)
one can easily get an extra 3\% deviation in $\alpha_s (M_Z)$.
Another possible source of uncertainty is the nonrelativistic
approximation heavily exploited in Ref. \cite{NRQCD}.
Although the authors write: ``it is clear that higher order relativistic
corrections ... would be completely invisible" I am not quite confident
that this is the case.

I should
add a few words about other groups. The Kyoto-Tsukuba group uses
the Wilson fermions to calculate the $\rho$ meson mass and the
charmonium energy levels \cite{Jap}. The $N_f$ extrapolation is also
carried out, as well as that in the sea-quark mass.  The result of the
group is $\alpha_s(M_Z) = 0.111\pm 0.005$. The conservative world
average quoted in the review talk \cite{CM} is
\begin{equation}
\alpha_s(M_Z)_{lattice} = 0.112\pm 0.007\, .
\end{equation}

\subsection{Problems with  $\alpha_s$ from
$R_\tau$}

My survey of the low-energy determinations of $\alpha_s$ would
not be complete without discussing the total inclusive widths of the
type $\tau\rightarrow \nu +$ hadrons or $\Upsilon\rightarrow$
hadrons. Let me focus on the first process simply because it is much
more often cited in the literature as a perfect source
of information on $\alpha_s$. One should keep in mind, however,
that conceptually theoretical analysis of both processes is essentially
identical.

As was noted above, the only low-energy ``determination"
of $\alpha_s$ which gravitates to the $Z$ cluster rather than to the
DIS one (i.e. yields a large value of $\alpha_s (M_Z)$) is that from
$R_\tau$.  Experimentally, $R_\tau$
is measured to a rather good accuracy,
$$
R_\tau = 3.65\pm 0.05\, .
$$
Braaten {\em et al.} suggested \cite{Pich}
to use this number in order to extract
the value of $\alpha_s (M_\tau )$, which, being evolved up to
$M_Z$, allegedly ensures high  accuracy in $\alpha_s (M_Z)$, see Fig.
1.

One can not deny apparently  appealing features of this proposal.
Indeed, in this problem the
state of the art analysis seems possible \cite{Pich}.
The perturbative corrections are known up to the third
order in $\alpha_s$.  Mass corrections can be readily accounted for
(they play little role, though). As far as the standard
non-perturbative corrections are concerned, formally they start from
the
gluon condensate. However, one  encounters here  a fortuitous
circumstance:
accidental cancellation of the coefficient in front of
$\langle {\cal O}_G\rangle M_\tau^{-4}$ in
the
leading
(one-loop) order. This cancellation of the gluon
condensate in $R_\tau$ is not backed up
by any symmetries, and as a matter of fact, does not persist
beyond  the leading order. Since the gluon condensate appears
only at the two-loop level, the corresponding contribution is
numerically small. The four-quark condensate appears
with the coefficient of the natural order of magnitude but  is
suppressed
by $M_\tau^{-6}$ and is also small numerically. The combined
effect of these two condensates is less than 2\% \cite{Pich}. If so,
$\alpha_s (M_\tau )$ could seemingly be determined from
the perturbative formula. Fitting the data for $R_\tau$ quoted above,
one
gets
$\alpha_s (M_\tau ) = 0.36\pm 0.03$ which results in a high
value of $\alpha_s (M_Z)$ displayed on Fig. 1.

The $R_\tau$ determination of
$\alpha_s$ was criticized in the literature more than once, see e.g.
\cite{R1,R4,R2}.
The aspect which was of greatest  concern so far, is the truncation of
the perturbative
series, and its factorially divergent structure which  shows
up already
in the third order since the coupling constant $\alpha_s (M_\tau )
\sim 0.3$.
Although this type of  uncertainty is definitely present
and affects the estimate of  possible errors of the prediction,
it seems unlikely that it alone is responsible
for  the value of $\alpha_s$ extracted from $R_\tau$ being too high.
The
renormalon chains are there irrespectively of whether we consider a
process in the Euclidean or Minkowski domains. I  think
that the most dangerous theoretical contribution, which as a rule is
not even mentioned, is specific to the processes with the essentially
Minkowskean
kinematics, like the hadronic width of $\tau$. Conceptually, it is
associated with the tail of the condensate
(non-perturbative) series. This tail, manifesting itself in exponential
terms
(invisible in the truncated condensate series) is small if viewed at
$M_\tau$
from the Euclidean side, but, unfortunately,  is decaying
much slower in the Minkowski domain and may be  quite noticeable
in the
Minkowski
analysis of $R_\tau$ carried out by Braaten {\em et al.} \cite{Pich}.
This
asymmetric nature of the condensate series is a general property
of QCD, it has nothing to do with the peculiarities of the
$\tau$ decay.

In  slightly different language the exponential contributions
not controllable by the truncated condensate series are called
{\em deviations from duality}. I can not dwell on this issue to the
extent I would like to, the more so that it  remains controversial.
Still, some facts may be considered established
\cite{pascos}.
 In particular, it
is known that although the deviations from duality are exponential
in the Euclidean and in the Minkowski sides, the fall off
is steep in the Euclidean domain and is much slower in the
Minkowski domain. Moreover, in the Minkowski domain,
the behavior of the exponential terms is {\em predicted
} to be modulated by oscillations. It is very difficult, if possible at all,
to make reliable predictions as to the specific law of the exponential
fall off of the deviations from duality, the character of modulations,
and the absolute value at given energy based on  present-day
QCD.  One has to resort to phenomenological information
(see Ref. \cite{pascos} for further details).

Fortunately, we do have such information relevant to the $\tau$
decays. To substantiate my point let me show you Figure 6
presenting what I call ``faked $\tau$ events". This is the plot of
the function
\begin{equation}
``R_\tau " /2 =
2\int_0^{M^2}
\frac{ds}{M^2}\left( 1-\frac{s}{M^2}\right)^2
 \left( 1+2\frac{s}{M^2}\right)\left[ R(s)\right]^{I=1}_{e^+e^-}
\label{fake}
\end{equation}
calculated by S. Eidelman using the
existing experimental data on  $\left[ R(s)\right]^{I=1}_{e^+e^-}$.
The difference between the faked   $\tau$ events and the actual
decay is as follows.
First, the spectral density used, refers only to the vector-vector
correlation, while in the $\tau$ decays one deals with the
vector-vector
plus the axial-axial. Second, the integral runs up to $M^2$ where
$M$ is a free parameter. In $R_\tau$ we have the very same integral
running up to $M_\tau$; needless to say that  $M_\tau$ is fixed
at its experimental value. The weight function in Eq. (\ref{fake}) is
the same as in the actual $\tau$ decays.  The omission of the axial
part of the spectral density plays no role in the aspect I want to
highlight here.

I have indicated only one typical error bar, since the errors are
100\%
correlated. They are statistically uncorrelated in $\left[
R(s)\right]^{I=1}_{e^+e^-}$ but are correlated in $R_\tau$ since
$R_\tau$
is the integral over all points. The whole curve can be shifted
up or down, by a standard deviation, as a whole,
but the uncertainty in the relative position of the points is
much smaller. This fact is of a paramount importance.

A single glance at Fig. 6 shows that the predicted oscillations do take
place.
It is clearly seen that at $M_\tau$ we are still {\em not} in the
regime
where one can use  asymptotic formulae --
i.e. truncated perturbative and condensate series. Indeed, at
$M_\tau$ we are at the middle of the second oscillation
which, presumably, must be followed by the third one
while the asymptotic formulae are absolutely smooth
and show no sign of oscillations. Of course,
one can always close one's eyes on the $M^2$ dependence and just
fit one point on the curve at $M=M_\tau$ thus producing a
``prediction"
for $\alpha_s (M_\tau )$.  The significance  of this ``prediction" is
close
to zero, however, since the theoretical contribution due
to the exponential terms is not under control. One can speculate that
the amplitude of the oscillations provides us with a measure of the
theoretical uncertainty.  If so, the error bar in $\alpha_s (M_\tau )$
must be  doubled compared to the number quoted above,
which means, in turn,  that $\tau$ determination of
$\alpha_s$ covers the whole interval from the low to the high
values, and is not informative.

Today we have no theoretical
method for
estimating the size of the exponential contributions. It is quite clear,
from  this  faked $R_\tau$ plot, that they are still
noticeable
at $M_\tau$ in the Minkowski domain. What can be said about
$\Upsilon\rightarrow$ hadrons? The leading condensate correction
in this process was found in Ref. \cite{oldV} and turns out to be
negligibly small. The condensate corrections, thus, are no menace to
extracting $\alpha_s$ from this total width. The perturbative
calculations
of the ratio $\Gamma (\Upsilon\rightarrow
\mbox{hadrons})/\Gamma (\Upsilon\rightarrow \mu^+\mu^- )$
were carried out in Ref. \cite{Up}, in the next-to-leading
approximation. The fit to $\Upsilon$,  $\Upsilon '$ and $\Upsilon ''$
yields \cite{ips} $\alpha_s (M_Z)_\Upsilon = 0.108\pm 0.005$,
where the error is dominated by the theoretical scale uncertainty
in the perturbative formula, and all three resonances give consistent
results. The experimental scatter is at the level 0.001 and can be
neglected.
 However,
{\em a priori} the analysis \cite{ips} is plagued by the same
shortcomings
as that of the $\tau$ decay: potentially uncontrollable exponential
terms in the theoretical prediction for the width. On the positive side,
the invariant mass of the hadronic state
here is essentially higher than in $\tau$, so that one may expect
that the oscillations die off.  On the other hand, the final
hadronic state in the $\Upsilon$ decay is obtained from the
gluon fragmentation, as opposed to the quark fragmentation in
$\tau$. It is known that deviations from duality are stronger in the
gluon world than in the quark one \cite{NSVZ}. Therefore, for the
time being I would
approach any results obtained from the total hadronic $\Upsilon$
width with extreme caution, even though they seemingly produce a
low value of $\alpha_s (M_Z)$ which I advocate here.

\section{If $\alpha_s(M_Z ) \approx 0.11$ then ...}

Summarizing the discussion in the previous part of the talk
I  conclude that
$\Lambda^{(4)}_{\overline{\rm
MS}}\approx
$ 200 MeV follows from all determinations other than those
at the $Z$ peak. I hope that you are convinced now that the genuine
value of
$\alpha_s(M_Z )$ is close to 0.11, not to 0.125.  Then we {\em have}
to ascribe the apparent excess of the
hadronic
decays
at the $Z$ peak  to new contributions unaccounted for in the
Standard Model.
First, I will consider immediate consequences for physics below 1
TeV and
then comment on implications for Grand Unification.

\subsection{New physics around the corner?}

Assuming that the experimental number at the $Z$ peak is
correct we are forced to accept that some contributions due to new
physics,
invisible at low energies, show up at the $Z$ peak.  Although at the
moment
nobody can definitely say what kind of new physics
will claim responsibility, it is curious to examine how
the existing popular scenarios  can cope with the
situation.

The most developed scenario is a supersymmetric generalization of
the Standard Model, the minimal supersymmetric standard
model (MSSM). This model introduces a superpartner to every
known
particle, plus two Higgs superfields. Moreover, one usually assumes a
very specific mechanism for
the supersymmetry breaking: a breaking in the invisible sector
through the gaugino condensation, transmitted to the visible sector
only through gravity \cite{Chams}. Then the masses of all gauginos
turn out to be the same at the Planck scale. The soft mass parameters
of all squarks and sleptons are set to be equal at this scale.  This
mechanism of the supersymmetry
breaking is so  popular among the SUSY model-builders that usually
they  do not differentiate between MSSM {\em per se} and the model
supplemented by the above {\em additional} assumptions. Below this
approach will be referred to as Constrained MSSM
(CMSSM).
I hasten to add that not a single experimental fact today points
out to the  existence of this particular mechanism of the
supersymmetry breaking.

The next step crucially depends on the additional information we
accept. Say, if we close our eyes on the $b\bar b$ excess
and $c\bar c$ shortage at all, and rely only on the $\alpha_s$
clash, the opportunities for speculations are very vast.
For instance, gluino and squarks in the 100 GeV ballpark
could push up a little all the hadronic channels uniformly ($u\bar u$,
$d\bar d $ and so on), so that the total excess of the hadronic width
can approach the desired 10 MeV \cite{E1}. Another solution which
might seem possible {\em a priori} is ultralight gluinos --
so light that they change the law of running of $\alpha_s$ in the
energy range of a few GeV \cite{Clav}.
It is clear, however, that
in this way it is impossible to explain the $\sim$10 MeV excess in
the
$b\bar b$ yield. Since this excess seems to be real, it is reasonable
to try  find a mechanism explaining simultaneously
the $b\bar b$ problem and the $\alpha_s$ problem.  As I mentioned
in  the beginning of the  talk, relaxing the mass relations of CMSSM
and arranging for a light stop and chargino helps do the job
\cite{Y}. The general feeling, however, is that getting the 10 MeV
excess in the $b\bar b$ yield is not an easy exercise: it requires
stretching the MSSM parameters   to their extremes, so that the
model is at the verge of  contradicting  the existing data
(or, perhaps, even beyond this line \cite{ELN}).
Similar
conclusions are achieved by other authors, as we heard today
in
Chankowski's talk \cite{E2}, although
particular details are somewhat different. The general feeling is that
the excess in the $b\bar b $ yield comes out too small in MSSM,
smaller than the experimental number.
Moreover,  all experts share the opinion that
this mechanism does not lead to any noticeable deviation in
the $c\bar c$ yield compared to the Standard Model (see e.g.
\cite{3,4,Y,E2}). Therefore, it looks like  we must turn to alternative
explanations.

In  absolute numbers, the $b\bar b$ excess amounts
to $11\pm 3$ MeV while the shortage in the $c\bar c$ channel is
$31\pm 13$ MeV \cite{Olch}. In the latter case, the relative
experimental uncertainty is significantly larger than in the $b\bar b$
channel, so it is not crazy to assume that the effect will just
evaporate with time.  This ``wait-and-see" attitude is perfectly
reasonable and is accepted by many. On the other hand, it is also
reasonable, and even
tempting, to assume that some $c\bar c$ shortage does take place.
The most attractive  possibility is the assumption that
the shortage in the $c\bar c$  channel is  the same as the
excess in the $b\bar b$ channel, say both are $\sim$ 14 MeV.
Combined with the information from $\alpha_s$ we can further
speculate that the yield of {\em both} up quarks, $u$ and $c$,
is suppressed by 14 MeV, while the yield of {\em all} down quarks,
$d$, $s$ and $b$ is enhanced by 14 MeV. Then the deviations from
the standard model in the yields of the first and second generation
quarks cancel  in the total hadronic width. If $t$ quark was light
enough
it would cancel the excess due to $b\bar b$. In the real world,
however, this excess remains uncompensated.

Could such a pattern emerge in a natural way? The answer is yes, at
least in principle, as is well-known to all those who played with
four generations \cite{NV1}. Indeed, let us assume that there
exists a new weak-isospin doublet of quarks (or
any other colored spinor fields), the masses of the {\em up} and
{\em down} components of this doublet are heavier than $M_Z$, but
not degenerate due to the spontaneous breaking of the
$ SU(2)_{weak}$ symmetry (let me call these components,
symbolically,
$T$ and $B$).  Then the diagram of Fig. 7 will produce
the pattern of deviations exactly as described above.  It goes without
saying that in supersymmetric generalizations the graph of Fig. 7 has
a tower
of accompanying counterpartners with sparticles in the loops.

$Z$ can proceed into two gluons only because of  axial coupling.
Moreover, due to the Landau-Pomeranchuk-Yang selection rules both
gluons then can not
be simultaneously on the mass shell; one of the gluon propagators is
contracted. The two-gluon intermediate state does not exist in this
mechanism,
a pleasant surprise by itself. As a matter of fact, the two-quark cut
shown on Fig. 7  dominates  since this contribution contains
logarithm of $m_T/m_B$ and is not suppressed by powers of
$1/m_{T,B}$. The quark-antiquark-gluon intermediate
state is suppressed by the second power of $1/m_{T,B}$ and can be
neglected if $2m_{T,B}\gg M_Z$.

The logarithmic nature of the cut shown on Fig. 7 is pretty obvious;
quantitatively one has \cite{log}
\begin{equation}
\delta\Gamma (Z \rightarrow b\bar b)
=2\left(\frac{\alpha_s}{\pi}\right)^2\, \left( \ln
\frac{m_B}{m_T}\right) \Gamma_A^{(0)}\, ,
\label{kuhn}
\end{equation}
where $\Gamma_A^{(0)}$ is the parton probability of the axial $Z$
decay into $b\bar b$ which amounts to $\sim
(1/10)\Gamma (Z) \sim 240 $ MeV.
Since $\delta\Gamma$ is due to interference between the amplitude
$Z\rightarrow 2g \rightarrow q\bar q$ (i.e. $I=0$) with the direct
$Zq\bar q$
coupling ($I=1$), the sign of $\delta\Gamma$ is different for
$q=u,c$ on one hand, and $q=d,s,b$ on the other. Thus,
the shortage of $c\bar c$ is automatically the same as the excess of
$b\bar b$.  The formula (\ref{kuhn}) is obtained with
the standard quarks, doublets with respect to $SU(2)_{weak}$
and triplets with respect to $SU(3)_{color}$. If the logarithm is of
order
one, the effect is way too small to be important, of the order of 1
MeV.

In principle, it is not so difficult to enhance it by an order of
magnitude by saying that $B$ and $T$ quarks are color octets or
weak isotriplets, and so on in the same vein. Moreover, in
supergeneralizations it may well happen that the tower of graphs
with sparticles in the loops add up coherently. This is not a problem.
The true problem is avoiding spoiling the successful predictions of
the standard model, say, the ratio of $M_Z/M_W$.  Experts say, that
new isodoublets contributing so significantly into
the imaginary part  will shift the ratio  $M_Z/M_W$
beyond the allowed limits \cite{Nmnogo}. The assertion
\cite{Nmnogo} refers, however, only to the standard heavy quarks,
replicas of the existing one. Whether it is still valid for color octets
or isotriplets, or a mixture of a little bit of everything, is not clear to
me at
the moment.

Of course, this mechanism is rather {\em ad hoc}; moreover, it goes
against the existing trend (or deep belief, if you wish) that any new
physics has to be associated exclusively with  MSSM. Well, at
this stage
it seems reasonable to keep our eyes open, making no ultimate
commitments based only on   theoretical prejudice or shaky
arguments. After all, in the past Nature surprised us more than once.

 It is worth noting that the induced $I=0$ vertex $Z\bar q q$
discussed above will also change all predictions for asymmetries and
sine-squared of the Weinberg angle.
If the induced $Z\bar b b$ coupling is purely axial the corresponding
change in the asymmetry ${\cal A}_b$ is too small to be important.
If, however, the ``new physics" contribution to $Z\bar b b$
involves predominantly the right-handed $b$ quarks (as may be the
case in some supersymmetric scenarios) then the induced $Z\bar b
b$ coupling noticeably changes the prediction for ${\cal A}_b$.
Adjusting the coupling in such a way as to reproduce $\delta\Gamma
(Z\rightarrow \bar b b)\sim $10 MeV we simultaneously
lower ${\cal A}_b$ by $\sim 8\%$ compared to the SM predictions
\cite{Hol}.
Remarkably, the value of ${\cal A}_b$
 detected at SLC is $\sim 8\%$ lower than the SM value ($\sim 2
\sigma$ effect)
 \cite{Olch}. The supersymmetric extension of the diagram of Fig. 7,
with two gluinos in the intermediate state, may produce the $I=0$
$Z\bar q_R q_R$ vertex provided that the right-handed squark
is essentially lighter than the left-handed one. It is curious that
if the diagram of Fig. 7 vanishes with switching off the SU(2)
symmetry breaking, its supergeneralization need not vanish in this
limit (although it vanishes, of course, with switching off
the supersymmetry breaking).  This is due to the $Z $ boson coupling
to the weak hypercharge.  Therefore, these graphs may bring in a
new scale. There are many open questions, though:

-- can the induced $Z\bar q_R q_R$ vertex contain  enhancement
factors of order 10  compared to the natural scale
$(\alpha_s/\pi)^2$?

-- if they do appear does this scenario go through the successful
predictions of the standard model?

Further detailed
analysis  is clearly in order.

 \subsection{Grand Unification}

Now I proceed to the second aspect of the small-versus-large
$\alpha_s$ problem -- Grand Unification.
It is difficult for me to go into details here
not only because this topic is vast and I have very little time left, but
also
because I am a newcomer in this field and still have to learn a lot.
I see experts in this audience  who will definitely
elaborate the point in their talks.

The idea of Grand
Unification is extremely attractive. I am aware of no other
explanation of the fact that the electric charge is quantized \cite{F1}.
At the same time I feel rather uneasy,   since Grand Unification
requires extrapolating the theory from the known range
of 100 GeV up to $10^{16}$ GeV, fourteen orders of magnitude.
I am a down-to-earth person and would like to
stay away from  speculations as to what is
going to happen with the theory on the way to $10^{16}$ GeV.
Still one most naive assumption -- essentially nothing happens --
is worth considering, say, for the purpose of getting
a proper reference point.

The plots presented below (Fig. 8), which I borrowed from
Langacker \cite{PL},
are well known. The first one, quite famous a couple
years ago, shows that in the Standard Model the evolution lines
of three coupling constants, $\alpha_1$, $\alpha_2$ and $\alpha_3$
do not intersect in one and the same point, so that the naive
straightforward
unification does not go through in SM. The second plot is meant to
be a triumph of the naive unification
within MSSM; it is designed to illustrate
that the three evolution lines perfectly intersect
in the naive Constrained MSSM (i.e. with no GUT threshold
corrections and/or non-renormalizable operators (NRO) \cite{NRO}
from the Planck
scale
physics added).  The only thing which remains to be added
is the value of $\alpha_s (M_Z)$ ensuring this intersection.
This value jiggles a little in different publications,
depending on how the SUSY threshold effects are treated, and
what corridor for the Weinberg angle is accepted,
but it jiggles between  0.125  and 0.128!  Alas, the ``triumph"
of the CMSSM Grand Unification is rather to be called a failure today.

Does this mean that the Grand Unification scenario
is ruled out? Certainly not. There exist three obvious ways
out:

(i) three gauge coupling constants need not intersect exactly at
one point if GUT threshold corrections and/or Planck scale
NRO's play a role;

(ii) the mechanism of SUSY breaking inherent to CMSSM can be
traded
for another mechanism where specific relations between the
sparticle masses appearing in CMSSM do not hold;

(iii) new physical phenomena can take place at an intermediate
scale, half way between the present-day 100 GeV and the GUT scale.

Let me comment on these three possibilities in turn. The first option
seems rather obvious and natural. One definitely expects some
effects
due to the fact that not all superheavy particles have exactly
the same mass. The size of the effects due to the GUT
scale thresholds (or NRO) has been investigated more than once (for
recent analyses see Ref. \cite{GUT, GUT1}).  In many instances they
come
out positive, i.e. lead to larger values of
$\alpha_s$, thus only aggravating the problem. Even if they can be
made negative the size of these corrections is typically not large
enough.  The interval of $\alpha_s (M_Z)$ emerging from the
supersymmetric GUT is believed to be \cite{Lang}  0.12 to 0.14,
with the typical prediction lying around 0.128.
I hasten to add, though, that other authors \cite{GUT1} seem to be
able to get down almost to 0.11.  Still, my feeling is that with
a natural  set of assumptions it is  difficult  to descend down to 0.11.
So what?

The strategy
which seems
more adequate to the present stage is as follows:
let us assume that $\alpha_s (M_Z)=0.11$ and find out what
can be said then about the GUT thresholds, and possible observable
implications at low energies. This strategy is pursued by
Lucas and Raby \cite{X}. Based on an overall analysis
of relevant operators in the SO(10)-based grand unification,
compatible with phenomenological pattern of masses and
angles, these authors find that $\alpha_s (M_Z)=0.11$
perfectly fits provided a certain relation between
the masses of the superheavy Higgses is accepted.
It is remarkable that this relation implies a significantly
lower prediction for the proton lifetime compared to the one
emerging in the naive constrained MSSM. The Lucas-Raby
result for the proton lifetime is much closer to the values accessible
to
experiment. This makes the issue of the
proton decay exciting again, extracting it from oblivion
it resided in during the last few years when people believed
that the theoretical expectation is so high there is no hope
of detecting  the proton decay experimentally in the
foreseeable future.

The second route to explore is relaxing some of the assumptions
of CMSSM, in particular the mechanism of the SUSY breaking.
If the SUSY breaking is transferred only through gravity
\cite{Chams}  the
masses of all gauginos are the same at the GUT scale. This
implies that at our scale gluino is $\sim$ 3 times heavier
than the wino. Other mechanisms of the SUSY breaking
were under discussion in the mid-eighties. In particular,
the one which seems promising  is
the instanton-generated dynamical breaking suggested
by Affleck {\it et al.} and Veneziano {\it et al.} \cite{SUSYB}.
Theoretically it
is an absolutely beautiful mechanism. Unfortunately,
no scheme which was phenomenologically nice and attractive was
found
in the eighties, and the whole idea has been in a dormant state since
then in connection with the rise of the gravity-induced
scheme of Chammseddine {\it et al.} mentioned above. Recently,
however,
the instanton ideas were revived by Dine {\em et al.} \cite{Dine},
who submerged into the search for a phenomenologically
successful and aesthetically attractive model with vigor. It is not
ruled out that
such a model will be found. Since this mechanism
is based on the SUSY breaking at a relatively low scale,
no relation between the gluino and wino masses emerges,
generally speaking. Anticipating the success of the search one
could ask the question: what happens with the prediction for
$\alpha_s (M_Z)$ from the naive Grand Unification (no GUT
thresholds and so on) if the relation between the gluino and wino
masses is relaxed?

The answer to this question was given recently by
Roszkowski and myself \cite{Ros}. As a matter of fact, the
result for $\alpha_s (M_Z)$ is essentially independent on all details
except the gluino and gaugino masses. It is not difficult to
get $\alpha_s (M_Z)\approx  0.11$, provided that the gluino is
relatively light (100 to 200 GeV) while the wino is
relatively heavy (600 GeV and heavier). Of course, the
precise number requires full calculations (which were performed)
but the tendency becomes immediately clear upon reflection
about what happens with the coefficients in the Gell-Mann-Low
functions
when one freezes out this or that particle. It is obvious that
making gluino light helps, as well as making the wino heavy.
Roughly, the ratio of the corresponding masses needed is 1/3,
i.e. inverse compared to what one gets in CMSSM (see Fig. 9).

Finally, the third option --
new physics, other than MSSM,  at an intermediate scale -- is
advocated, for instance, in
Refs.
\cite{X1,X2}.
As you could see from the previous section, I absolutely do not rule
out and, on the contrary, to an extent expect the advent of ``new"
new
physics in the great desert predicted by MSSM in the energy range
from $\sim 10^2$ to $\sim 10^{16}$ GeV.
This
is a viable option, but it is very difficult
to pursue this scenario, since we have very little evidence
to assume something definite, and, hence,
very little predictive power. The same refers to non-renormalizable
operators coming from the Planck scale -- string physics. They
definitely could enter the game
\cite{Faraggi},
but since very little is known about physics at this scale, and too
much
depends on pure speculation, I prefer to refrain
from going into this topic, leaving the field
completely open to experts.

Summarizing, the issue of Grand Unification has made a full turn
now;
we are back to square one. While the most naive version, with
no new effects (apart from superpartners) in the interval from 100
GeV up to
the GUT scale,
produces too high a value of $\alpha_s (M_Z)$ refinements
can, in principle,  bring this value down to 0.11. The question is
what refinements are to be done. It seems that additional
information
(or divine guidance) is needed in order to answer the question.

\section{Conclusions}

I still remember that in 1970, before the QCD era and before the
Standard Model, some people discussed $K_L-K_S$ mass difference
in the most naive manner, calculating diagrams with free quarks,
and concluding that the small value of this mass difference can be
explained
in no other way than a new particle, charmed quark, compensating
for a large contribution coming from the virtual $u$ quark.
Many  were very skeptical, since
 these calculations were perceived as far too naive \cite{Ioffe}, and,
moreover,
this
was just one number, one contradiction with an analysis which
did not smell too clean. So, they preferred to wait. We all know now
who was right and who was wrong. I think that now Nature gives us
a very similar sign, perhaps the only one we can get from it
 with the existing machines. It would be unforgivable to loose this
chance.
Since 1974, we have waited for a miracle, for the advent of new
physics. And
now,
when it seems already with us, we must keep our eyes open to it.
If we are optimistic enough,  we should say that new physics
is already discovered and try to extract maximal information
from this fact. For those who are more cautious I can only repeat
that (i) the $\alpha_s$ testimony can not be overruled, and (ii) we
definitely have now at our disposal
{\em three or four} phenomena where 2 to 3$\sigma$ deviations
from SM are detected. These observations are uncorrelated
experimentally, therefore the probability of statistical fluctuations
must be multiplied, which leaves us with a pretty
improbable interpretation, at the level of $10^{-4}$ or less,
unless we invoke new physics.

Finally, the minimal
lesson refers to the QCD practitioners. It became fashionable
in many works devoted to the low-energy hadronic physics, to use,
as the most precise value of $\alpha_s$ the one stemming
from the determination at the $Z$ peak. The 13\% shift in $\alpha_s$
at
high energies in many instances produces
quite a dramatic effect at low energies. I would like to urge not
to follow this fashion blindly. The determination from the $Z$ peak
is {\em not} the most precise one, for the reasons I have explained.
The low-energy calculations should use the value of $\alpha_s$
extracted from the low-energy processes, which is
inconsistent with that determined at the $Z$.

\vspace{1cm}

{\bf Acknowledgments}\hspace{0.7cm} I am grateful to S.  Eidelman,
A.
Kataev, L. Roszkowski, A. Vainshtein and M. Voloshin for extremely
valuable comments. S. Eidelman  provided me with the numerical
data on which Figures 2 and 6 are based. I would like to thank D.
Harris for communicating to me relevant CCFR experimental data.

\newpage

\vspace{0.5cm}
\epsfxsize=100truemm
\centerline{\epsfbox{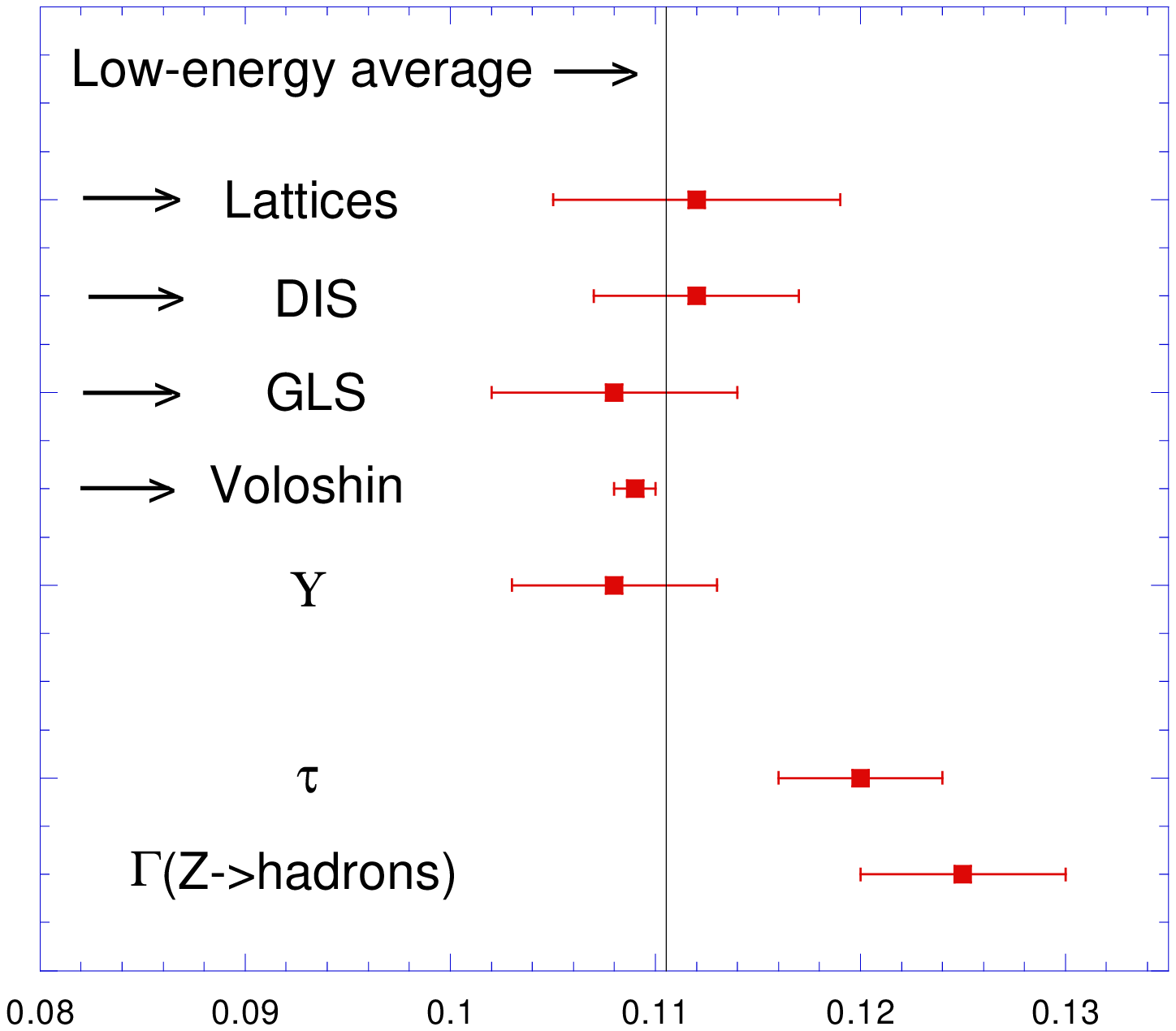}}

\vspace{1cm}

Fig. 1. Experimental data on $\alpha_s (M_Z)$. (Udapted from Ref.
\cite{Bethke}). The vertical line is a naive average of four low-energy
points marked with the arrows.

\newpage

\vspace{0.5cm}
\epsfxsize=130truemm
\centerline{\epsfbox{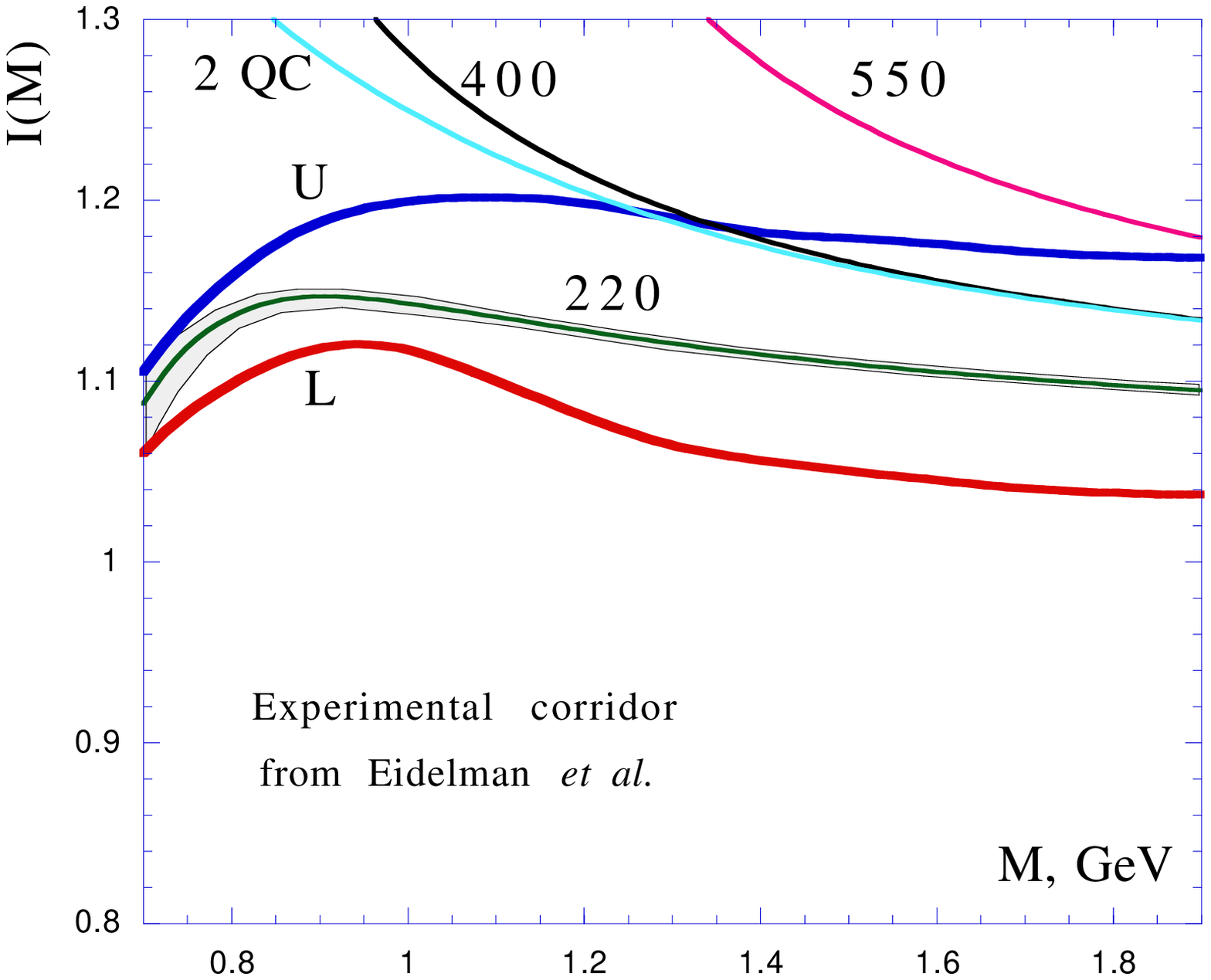}}

\vspace{1cm}

Fig. 2. The QCD sum rule in the $\rho$ meson channel
for different values of the scale parameter
$\Lambda^{(3)}_{\overline{\rm
MS}}$. The curves marked by ``U" and ``L" are the upper and lower
sides of the experimental corridor (see text and Ref. \cite{EKV}). The
curve marked by ``2 QC" corresponds to
$\Lambda^{(3)}_{\overline{\rm
MS}} = 400$ MeV and the doubled value of the quark condensate.

\newpage

Fig. 3.
Determining the strong coupling constant from deep inelastic
scattering  (from Ref. \cite{VM}).

Fig. 3a.
Next-to-leading order QCD fit to the deuterium structure
function measured by SLAC and BCDMS in the central $x$ domain.
The solid line is the result of the fit;
the dashed line vizualizes the $Q^2$ evolution without the higher
twist effects (the target mass corrections are included, however).

Fig. 3b.
The logarithmic derivative $d\ln  F_2(x,Q^2)/d\ln Q^2$ at a high
value of $Q^2$.
The solid curve corresponds to $\alpha_s (M_Z) =0.113$. The dashed
curves correspond to 0.123 and 0.103.

\vspace{1cm}

Fig. 4. The variation of $\alpha_s(M_Z)$ as a function of a scale
parameter $k$ (from Ref. \cite{VM}).

\vspace{1cm}

\vspace{0.5cm}
\epsfxsize=100truemm
\centerline{\epsfbox{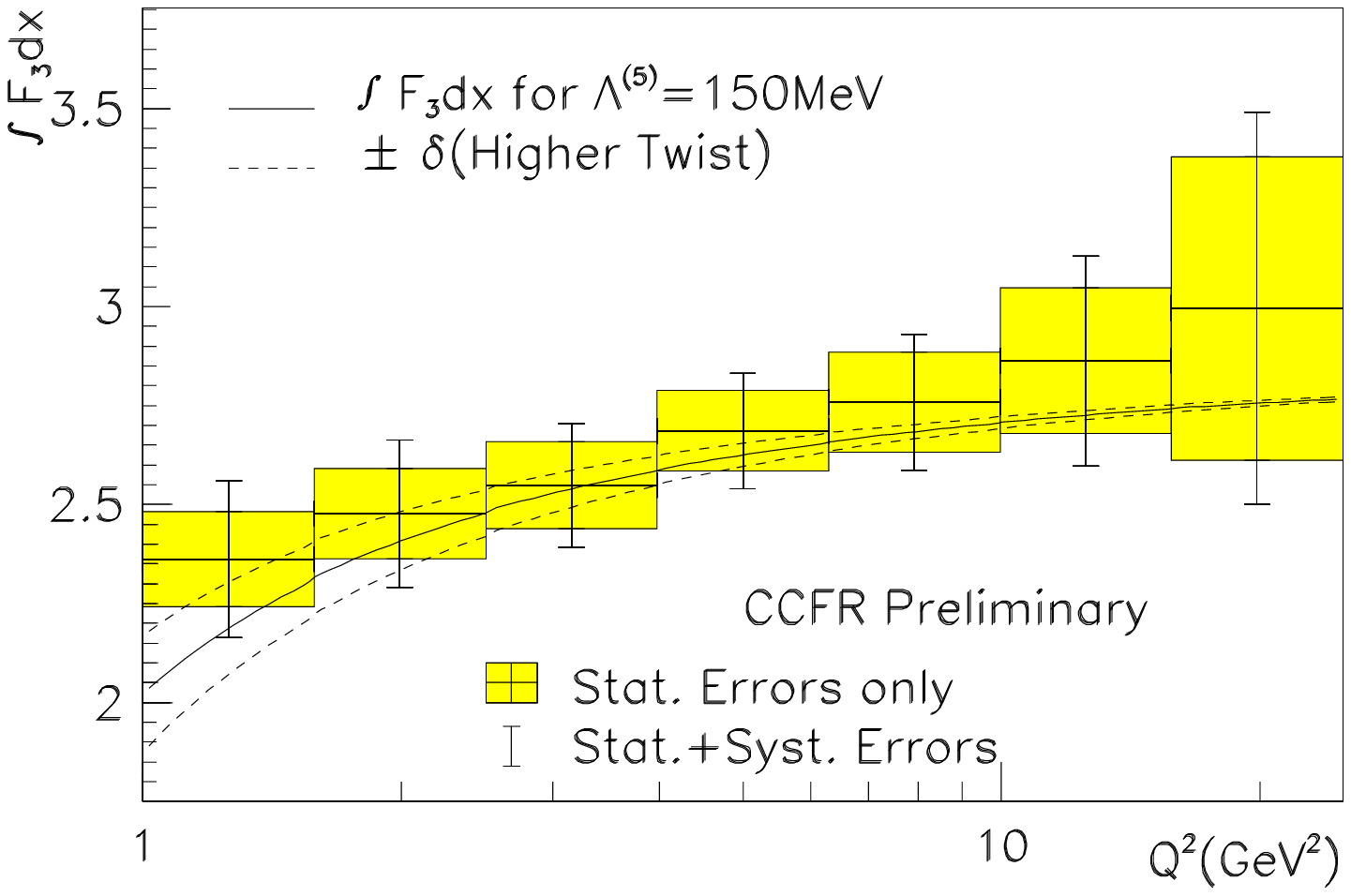}}

Fig. 5. The QCD fit to the Gross-Llewellyn Smith sum rule.
The solid
line is the result of the fit corresponding to
$\alpha_s(M_Z) = 0.112$  (from Ref. \cite{Harris}).

\newpage

\vspace{0.5cm}
\epsfxsize=120truemm
\centerline{\epsfbox{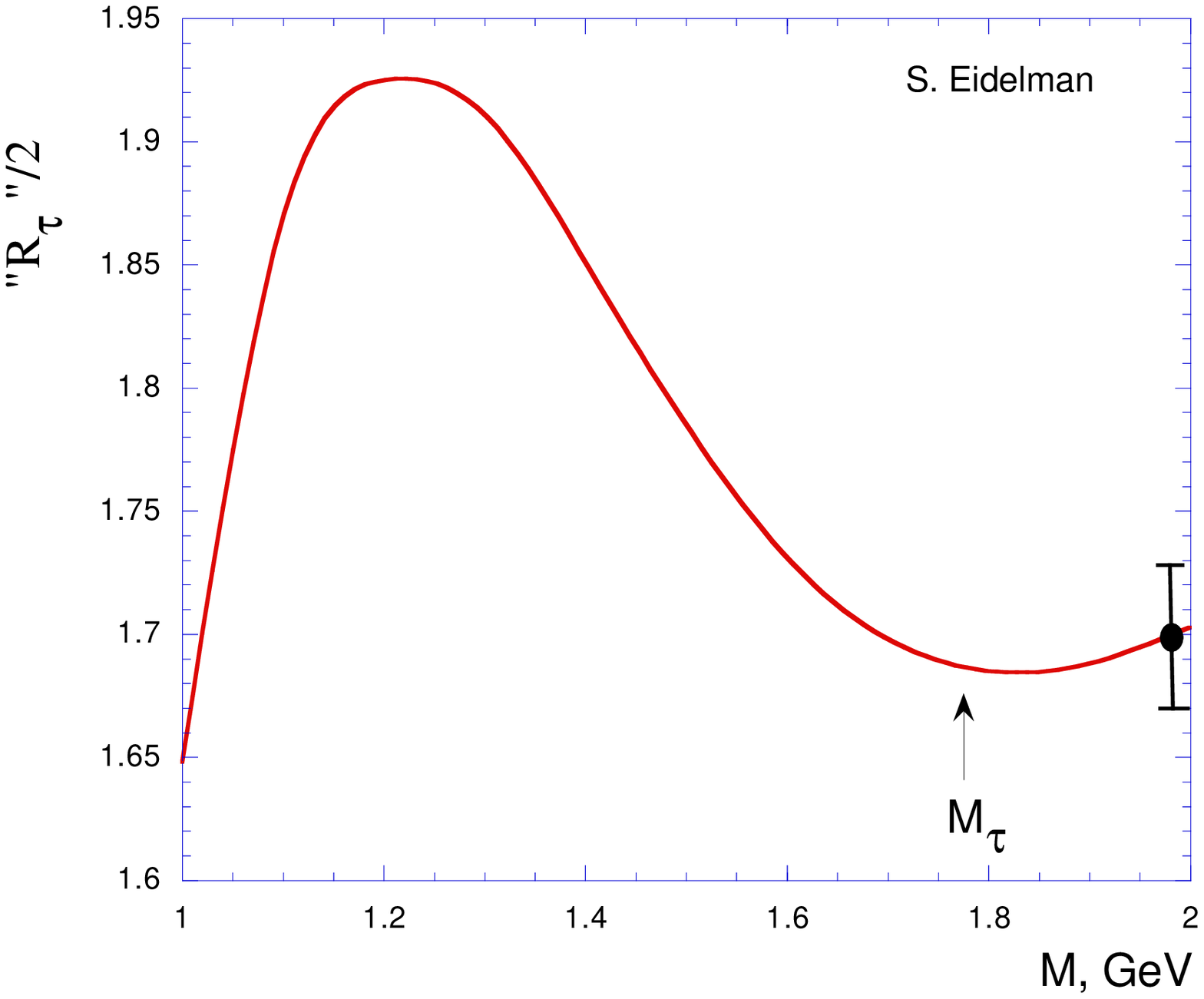}}

\vspace{1cm}

Fig. 6. Eidelman's fake $\tau$ events versus the faked $\tau$ mass
(see text). The curve can be shifted as a whole up or down
within the limits indicated by the error bar on the right-hand side.

\newpage

\vspace{0.5cm}
\epsfxsize=60truemm
\centerline{\epsfbox{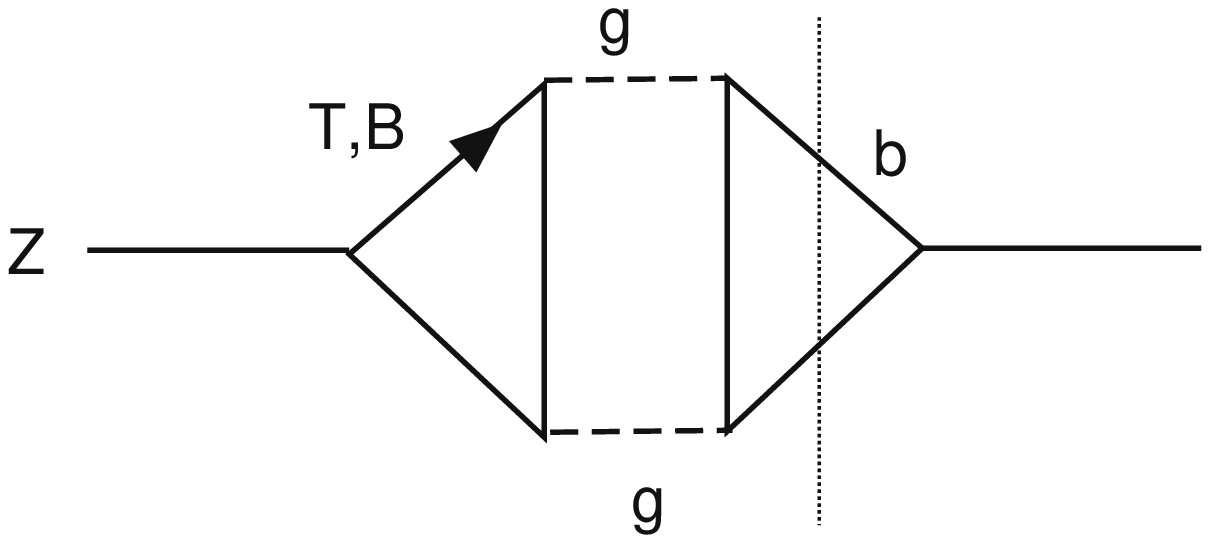}}

\vspace{1cm}

Fig. 7. The two-loop graph giving rise to $\delta\Gamma
(Z\rightarrow \bar b b)$. The $\bar b b$ cut is indicated by the
dotted line.

\vspace{1cm}

Fig. 8. The evolution of the coupling constants within the Standard
Model and MSSM (from Ref. \cite{PL}).

\newpage

\vspace{0.5cm}
\epsfxsize=120truemm
\centerline{\epsfbox{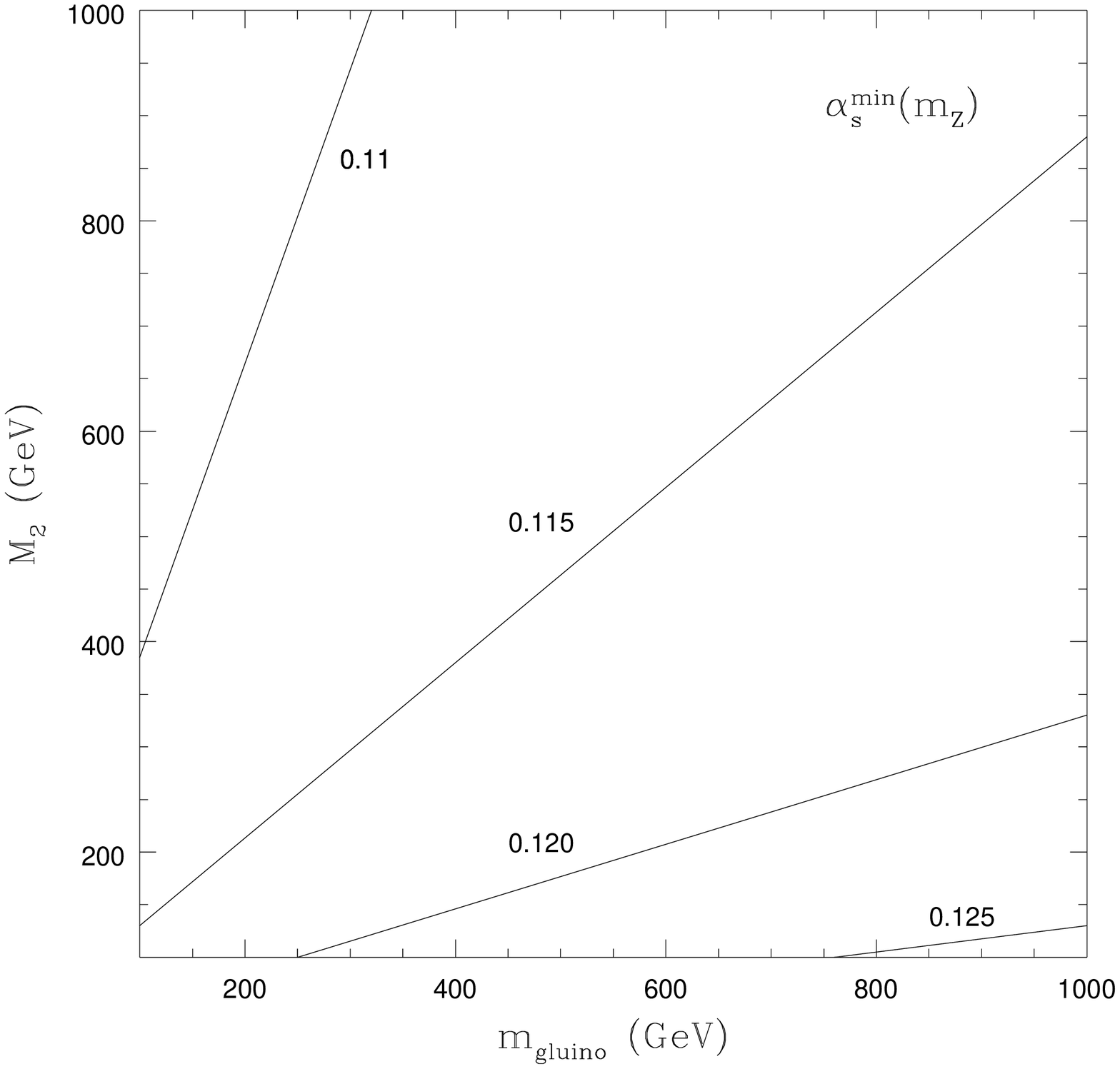}}

\vspace{1cm}

Fig. 9. Contours of  constant $\alpha_s (M_Z)$ in the plane $
(m_{\tilde{g}}, M_2)$ (from Ref. \cite{Ros}). All other parameters are
chosen in such a way as to minimize $\alpha_s (M_Z)$. The upper left
corner of the plot presents a phenomenologically
acceptable domain.


\begin{thebibliography}{99}
\bibitem{F1}
A weaker requirement of cancellation of all triangle anomalies
in the Standard Model also
leads to the electric charge quantization. This condition, in turn,
can be considered as a natural consequence of Grand Unification.
\end{thebibliography}
\end{document}